\documentclass[11pt,preprint]{aastex}
\usepackage{amsmath}

\def\ltsima{$\; \buildrel < \over \sim \;$}
\def\simlt{\lower.5ex\hbox{\ltsima}} 
\def\gtsima{$\; \buildrel > \over \sim \;$}
\def\simgt{\lower.5ex\hbox{\gtsima}} 
\def\arcsec{\hbox{$^{\prime\prime}$}}
\def\deg{\hbox{$^\circ$}}

\slugcomment{submitted to ApJ}
\shorttitle{Hotspots of Cyg~A}
\shortauthors{Stawarz et al.}

\begin{document}

\title{The Electron Energy Distribution in the Hotspots of Cygnus~A :\\ Filling the Gap with the {\it Spitzer Space Telescope}}

\author{\L . Stawarz$^{1, 3}$, C.~C. Cheung$^{1, 4}$, D.~E. Harris$^2$, and M. Ostrowski$^3$}
\affil{$^1$Kavli Institute for Particle Astrophysics and Cosmology, Stanford University, Stanford CA 94305\\ 
$^2$Harvard-Smithsonian Center for Astrophysics, 60 Garden St., Cambridge MA 02138\\
$^3$Astronomical Observatory, Jagiellonian University, ul. Orla 171, 30-244 Krak\'ow, Poland\\
$^4$Jansky Postdoctoral Fellow of the National Radio Astronomy Observatory\\
E-mails: \texttt{stawarz@slac.stanford.edu} , \texttt{teddy3c@stanford.edu} , \texttt{harris@head-cfa.harvard.edu} , \texttt{mio@oa.uj.edu.pl}}

\begin{abstract}

Here we present {\it Spitzer Space Telescope} imaging of Cyg~A with the Infrared Array Camera at $4.5$\,$\mu$m and $8.0$\,$\mu$m, resulting in the detection of the high-energy tails or cut-offs in the synchrotron spectra for all four hotspots of this archetype radio galaxy. When combined with the other data collected (and re-analyzed) from the literature, our observations allow for detailed modeling of the broad-band (radio-to-X-ray) emission for the brightest hotspots A and D. We confirm that the X-ray flux detected previously from these features is consistent with the synchrotron self-Compton radiation for the magnetic field intensity $B \approx 170$\,$\mu$G in hotspot A, and $B \approx 270$\,$\mu$G in hotspot D. We also find that the energy density of the emitting electrons is most likely larger by a factor of a few than the energy density of the hotspots' magnetic field. We construct energy spectra of the radiating ultrarelativistic electrons. We find that for both hotspots A and D these spectra are consistent with a broken power-law extending from at least $100$\,MeV up to $\sim 100$\,GeV, and that the spectral break corresponds almost exactly to the proton rest energy of $\sim 1$\,GeV. We argue that the shape of the electron continuum most likely reflects two different regimes of the electron acceleration process taking place at mildly relativistic shocks, rather than resulting from radiative cooling and/or absorption effects. In this picture the protons' inertia defines the critical energy for the hotspot electrons above which Fermi-type acceleration processes may play a major role, but below which the operating acceleration mechanism has to be of a different type. At energies $\gtrsim 100$\,GeV, the electron spectra cut-off/steepen again, most likely as a result of spectral aging due to radiative loss effects. We discuss several implications of the presented analysis for the physics of extragalactic jets.

\end{abstract}


\section{Introduction}

Cygnus~A is one of the best studied examples of Fanaroff-Riley type II radio galaxies \citep[`classical doubles',][]{fan74}, being uniquely observed at different scales and wavelengths thanks to its proximity\footnote{At the redshift of $z = 0.0562$ \citep{sto94} the luminosity distance to Cyg~A is $d_{\rm L} = 248.2$\,Mpc, and the conversion scale is $1\arcsec = 1.079$\,kpc for the modern cosmology with $\Omega_{\rm M} = 0.27$, $\Omega_{\Lambda} = 0.73$, and $H_0 = 71$\,km\,s$^{-1}$\,Mpc$^{-1}$.} and high total radio power \citep{car96,ch96}. The distinctive hotspots near the extremities of the radio structure are roughly $70$\,kpc from the nucleus in the North-West \citep[A and B on the jet side; designation by][]{har74} and the South-East (D and E on the counter-jet side) directions, and are the brightest known radio hotspots. They have been detected across the entire accessible radio band from $\lesssim 100$\,MHz up to $\gtrsim 100$\,GHz \citep[e.g.,][]{car91,wri04,laz06}. These features are understood as the terminal regions of relativistic jets, where bulk kinetic power transported by the outflows from the active center is converted at the strong shock (formed due to the interaction of the jet with the ambient gaseous medium) to the internal energy of the jet plasma \citep{bla74,sch74}. The nonthermal hotspot emission is believed to originate mostly in the downstream region near the reverse shock front, just before the shocked jet plasma is turned aside by the ambient medium and forced to flow into the extended radio lobe \citep[see a general review by][]{beg84}.

Besides this general picture, details of both hotspots' macroscopic structure and microscopic processes taking place therein are not well explored, mostly because crucial parameters of the jets (like the jet composition or bulk velocity on large scales) are still hardly known. Numerical hydrodynamical studies indicate a complex and time-dependent morphology of the jet termination regions, with an important role played by turbulent vortices and small-scale oblique shocks formed at the edges \citep[e.g.,][]{mar97,sch02,sax02,miz04}. Overall hotspots' structures like the ones observed in classical doubles can be however relatively well reproduced in these simulations with the jet outflow assumed to be dynamically dominated by cold particles (either electron-proton or pure electron-positron pairs). Much worse is the situation regarding microscopic processes leading to the conversion of the jet kinetic energy to the internal energy of the jet plasma (both particles and the magnetic field) at the reverse shock front, involving particle acceleration and possible energy equilibration among different plasma species. In particular, although the diffusive (1st order Fermi) shock acceleration was widely assumed in the past as the process of choice for generation of the broad-band non-thermal electron energy distribution within the hotspots \citep[see][]{hea87,mei89}, the most recent studies indicate serious problems with such a picture. These are mostly due to the expected (at least mildly) relativistic velocities of the jets in FR~II sources on large scales and oblique magnetic field configuration, precluding particles from undergoing 1st-order acceleration process \citep[see, e.g.,][discussing different aspects of this issue]{beg90,hos92,nie04,nie06,lem06}.

Some of the problems mentioned above regarding understanding of the hotspots (and thus of extragalactic jets) can be overcome by putting observational constraints on the theoretical models by means of accessing, and detailed modeling of, the broad-band hotspots' spectra. Cyg~A is of primary importance in this context, since its bright hotspots can be detected and resolved at different wavelengths. And indeed, {\it ROSAT} observations reported by \citet{har94} revealed relatively intense X-ray emission of hotspots A and D, being consistent with the synchrotron self-Compton (`SSC') model for the hotspots' magnetic field of $\sim (100-300)$\,$\mu$G in energy equipartition with the emitting electrons. These important findings --- in fact the very first observational suggestion for the minimum energy condition fulfilled in the case of extragalactic jets --- were confirmed recently by the {\it Chandra X-ray Observatory} \citep{wil00}. Still, the insufficiently sampled synchrotron continuum of the analyzed features (unknown minimum and maximum synchrotron frequencies in particular) made these conclusions not fully robust \citep[see, e.g.,][]{kin04}. Hence an observational effort was made to cover the whole synchrotron spectra in Cyg~A hotspots, both at their low-frequency \citep{mux88,car91,laz06} and high-frequency \citep{mei97,nil97,car99} ends, as well as to construct a large sample of other hotspots with multiwavelength data available \citep[see][and references therein]{bru03,har04,kat05,che05}. This is also crucial for understanding particle acceleration taking place at relativistic shocks.

Here we present {\it Spitzer Space Telescope} imaging of Cyg~A with IRAC at $4.5$\,$\mu$m and $8.0$\,$\mu$m, resulting in detection of the high-energy tails or cut-offs in the synchrotron spectra for all four hotspots of this archetype radio galaxy. This, together with the archival data collected from the literature (see \S~2), allows us to construct the broad-band spectral energy distributions for the discussed features, to model them in terms of synchrotron and SSC processes (\S~3), and thus to put meaningful constraints on the physical processes taking place at the terminal shocks of relativistic jets (\S~4).

\section{The Data}

\subsection{{\it Spitzer Space Telescope} Observations}

We observed Cygnus~A with the Infrared Array Camera \citep[IRAC;][]{faz04} aboard {\it Spitzer} on 21 October 2005, and the resulting image is shown in Figure~1. Individual $30$ second exposures were obtained simultaneously at $4.5$ and $8.0$\,$\mu$m over a $36$ position Reuleaux pattern ($1/4$\,th sub-pixel sampling). This dither pattern results in even coverage of a central $\sim 4 \times 4$\,arcmin$^{2}$ field which safely encloses the full extent ($\sim 140\arcsec$) of the Cyg~A radio source. The total field coverage is $\sim 6 \times 6$\,arcmin$^{2}$ with the exposure decreasing by about $\times 1/4$\,th of the total in the outer $\sim 1$\,arcmin wide edges. We repeated the dither $12$ times for a total integration time of $3.6$\,hrs in the central portion of the field. The $3 \sigma$ point source detection limits in the central field are $0.4$ ($4.5$\,$\mu$m) and $3.0$ ($8.0$\,$\mu$m) $\mu$Jy (Spitzer Observer's Manual). 

The pipeline processed basic calibrated data (BCD) files of the individual exposures were obtained from the {\it Spitzer} Science Center and combined with the MOPEX package \citep{mak05} resampling to $1/3$\,rd of the native pixel sizes' $1.22\arcsec$ (there are $< 1 \%$ distortions across the FOV and between detectors which we ignore). The hotspots were expected to be faint in the mid-IR so we required precise image registration with the other wavelength data (\S~2.2, \S~2.3). To facilitate this registration, the frametime of $30$ seconds was chosen to be long enough to minimize the noise per exposure while also short enough to ensure an unsaturated image of Cyg~A in at least one IRAC band ($4.5$\,$\mu$m). Based on the radio core and position of the galaxy in the $4.5$\,$\mu$m image, the alignment is better than $1/2$ of the resampled IRAC pixel, i.e. $< 0.2\arcsec$. Cyg~A was only slightly saturated in the $8.0$\,$\mu$m image but inspection of field sources appearing in both IRAC channels show the registration to be similarly good. 

We performed aperture photometry using $4.88\arcsec$ ($12$ resampled pixels) diameter circles centered on the hotspots and identical adjacent apertures for background determination (Figure~2). The aperture correction for this aperture is $22.1\%$ and $57.1\%$ at $4.5$ and $8.0$\,$\mu$m, respectively \citep{lai05}. Additional extinction corrections of $2.67\%$ ($4.5$\,$\mu$m) and $2.31\%$ ($8.0$\,$\mu$m) were estimated from the B-band extinction of $1.644$ mag \citep{sch98} extrapolated to our bands assuming $A_{\rm 4.5 \, \mu m} / A_{\rm B} = 0.0174$ and $A_{\rm 8.0 \, \mu m}/A_{\rm B} = 0.0151$ \citep{rie85}. 

\subsection{Radio Data}

There is a wealth of information on the Cyg~A hotspots in the literature. We focused on compiling radio interferometry data with sufficient resolution and sensitivity to detect and separate the differing contributions of the primary (i.e., weaker and compact) hotspots B and E from the secondary (i.e., brighter and extended) hotspots A and D. \citet{wri04} reported such a study utilizing the {\it Very Large Array} ($5$ and $15$\,GHz) and the {\it BIMA} Array ($87$ and $230$\,GHz) data and we have used these measurements. These authors used $10 \%$ calibration uncertainties for the {\it VLA} data and $12 \%$ for the {\it BIMA} ones. The {\it BIMA} $230$\,GHz images supersede previous non-interferometric mm/submm-wave measurements \citep[e.g.,][]{rob98}. In particular, the previous data were of lower resolution and did not resolve the primary from the secondary hotspots so did not suit our purposes. Additionally, the integrated fluxes of the SE and NW hotspots in the {\it SCUBA} data \citep{rob98} are systematically a factor of $\sim$two larger than the equivalent {\it BIMA} fluxes. We do not have an explanation for the discrepancy, but simply comment that the {\it BIMA} data are more consistent with the lower frequency ($5-87$\,GHz) extrapolation (Figures~3-6). 

At frequencies below $1$\,GHz, it was noted by \citet{car91} that the spectra of the secondary hotspots A and D flatten. The reason for such could be intrinsic flattening or cut-offs in the hotspots' electron energy distribution at low energies \citep[as favored by][]{car91}, or an absorption effect, either synchrotron self-absorption or free-free absorption by external (foreground) thermal medium. The most recent high-resolution analysis by \citet{laz06} confirmed the presence of low-frequency curvature, indicating in addition that some kind of absorption (being pronounced below $100$\,MHz) seems to be present in hotspot A, but not necessarily in hotspot D. For this reason, we restrict our discussion to data at frequencies $> 100$\,MHz for which absorption effects are expected to be minor (if present at all), and the interferometers' beams are small enough to extract fluxes from the same hotspot volumes as specified in the previous section. In particular, we analyzed the $327$\,MHz map ({\it VLA $+$ Pie Town Link Connected Interferometer}) with a $2.5\arcsec$ beam kindly provided by T.~J.~W. Lazio, and measured the fluxes for hotspots A, B and D as listed in Table~1. Similarly, we utilized a $1.34$\,GHz {\it VLA} image \citep[provided by C.~L. Carilli, and discussed previously in][]{car91}.

\subsection{Previous Optical and Infrared Observations}

The optical and infrared observations of Cyg~A are hampered by the fact that this source is at low Galactic latitude ($b = 5\deg$). In addition, bright stars and other optical foreground/background objects are present in the close vicinities of the Cyg~A hotspots, making it difficult to precisely determine the intensity of the relatively weak hotspot emission. Because of this, \citet{mei97}, using the {\it United Kingdom Infrared Telescope}, were able to provide only upper limits (see Table~1), for example $< 50 $\,$\mu$Jy and $< 5$\,$\mu$Jy for hotspots A and D, respectively, at $0.452$\,$\mu$m. Similarly, studies with the {\it Infrared Space Observatory} reported by \citet{car99} resulted in upper limits for the hotspots' emission at $12$\,$\mu$m, indicating however some attenuation of the hotspots' synchrotron fluxes at infrared wavelengths when compared with the extrapolation of their radio continua. Here we adopt conservatively the $3 \sigma$ upper limits from {\it ISO} observations, namely $285$\,$\mu$Jy and $255$\,$\mu$Jy for hotspots A and D, respectively (see Table~1). We also note that these {\it ISO} limits are above the IRAC extrapolations for all the discussed features. \citet{nil97} performed careful analysis of the archival {\it Hubble Space Telescope} data for Cyg~A, and found non-thermal optical emission associated with hotspots B and D. The work by \citet{nil97} remained almost completely unnoticed, but here we include their measurements (converted from magnitudes to the flux units as given in Table~1), arguing that they match the broad-band spectra of both hotspots. In particular, we show that these optical fluxes are consistent with the low-frequency segment of the SSC component.

\subsection{X-ray Data}

The X-ray fluxes in this paper are taken from \citet{wri04}, who re-analyzed the archival {\it Chandra} data for Cyg~A reported previously by \citet{wil00}. In particular, we adopt the results of their modeling involving an additional intrinsic absorbing column density for all the X-ray detected hotspots A, B, and D, being in a range $N_{\rm H}^{\rm int} = (3.6 - 4) \times 10^{21}$\,cm$^{-2}$, in addition to the canonical Galactic absorbing column toward Cyg~A source $N_{\rm H} = 3.06 \times 10^{21}$\,cm$^{-2}$ (see Table~1). The additional X-ray absorption leads to slightly larger intrinsic X-ray fluxes and slightly steeper $0.5-6$\,keV intrinsic continua for the discussed features, than obtained previously.

\section{Modeling the Broad-Band Emission}

Figures~3 and 4 show the broad-band spectra and spectral energy distributions, respectively, for all the Cyg~A hotspots. All of these spectra at low-frequencies, $\nu \leq 1$\,GHz, can be fitted by flat power-laws with spectral indices $\alpha < 0.5$ (where $S_{\nu} \propto \nu^{-\alpha}$), except for the weakest hotspot E for which the $327$\,MHz flux could not be determined. At higher frequencies, $\nu > 1$\,GHz, the spectra steepen up to $\alpha \gtrsim 1$, and cut-off or break again around $\gtrsim 10$\,$\mu$m as indicated by the {\it Spitzer} detections (and also {\it ISO} upper limits). These are then the synchrotron components of the Cyg~A hotspots. The optical $\lesssim 1$\,$\mu$m emission detected from hotspots D and B is stronger than expected from extrapolation of the synchrotron continua to higher frequencies in the case of hotspot D, but not necessarily in the case of hotspot B. As mentioned above, this emission is however consistent (in both cases) with the low-frequency segment of the SSC component detected at keV photon energies. Moreover, the optical-to-X-ray power-law slopes $\alpha_{\rm O-X} \gtrsim 0.6$ are in reasonable agreement with the X-ray spectral indices found in the previous studies. We note that there is no other hotspot for which SSC emission has been detected at optical frequencies, with the possible exception of the radio galaxy 3C~196 \citep{har01}, for which however no X-ray or infrared data are available.

In general, hotspots A and B on the jet side are much brighter in the infrared than hotspots D and E on the counterjet side. This is most clearly manifested for the secondaries A and D, which are comparably bright in radio (especially at low frequencies) and in X-rays, but which differ more than an order of magnitude in flux at $4$\,$\mu$m. The infrared continuum of hotspot A seems to be flatter than that of hotspot D, while at $(1-100)$\,GHz frequencies the opposite behavior can be noted. Possible reasons for these are discussed later on. Meanwhile, we concentrate on modeling of the multiwavelength emission of the bright hotspots D and A.

\subsection{Hotspot D and A: SSC Emission and Energy Equipartition}

Comparing the equipartition value of the magnetic field intensity deduced from the observed synchrotron emission of the non-thermal magnetized plasma, $B_{\rm eq}$, and the value suggested by the analysis of the accompanying SSC radiation sampled by a single (e.g., X-ray) flux density, $B_{\rm ic}$, is never a straightforward procedure. One of the problems here regards evaluation of the minimum energy condition, for which the crucial issue is the detailed shape of the synchrotron continuum, being poorly constrained by observations in some segments. Thus, more or less arbitrary assumptions have to be invoked in this respect, which may influence the resulting value of $B_{\rm eq}$. The other problem is connected with the very nature of the SSC emission, namely with the fact that in inverse-Comptonization of the broad-band target photon population by the broad-band electron population there are particles and target photons with different energies which contribute to the monochromatic SSC flux at some particular inverse-Compton photon energy. As a result, single-flux-estimated $B_{\rm ic}$ is in fact a function of the partly assumed synchrotron spectral shape (or, equivalently, of the electron energy distribution). However, relatively well-sampled synchrotron continua of hotspots D and A, together with the SSC component detected at two (optical and X-ray) frequencies in the former case, allow us to minimize the effects of the assumptions.

We model the synchrotron component of hotspot D as
\begin{equation}
S_{\nu}^{\rm syn} \propto \left\{ \begin{array}{ccc} \nu^{-\alpha_1} & {\rm for} & \nu_{\rm min} < \nu < \nu_{\rm cr} \\
\nu^{-\alpha_2} \, \exp\left(-\nu/\nu_{\rm max}\right) & {\rm for} & \nu > \nu_{\rm cr}
\end{array} \right. \quad ,
\end{equation}
\noindent
and $S_{\nu}^{\rm syn} \propto \nu^{1/3}$ for $\nu < \nu_{\rm min}$. Here $\alpha_1 = 0.21$, $\alpha_2 = 1.1$, $\nu_{\rm cr} = 3$\,GHz, and $\nu_{\rm max} = 0.9 \times 10^{13}$\,Hz are the spectral indices and critical (break) and maximum frequencies implied by the radio-to-infrared data as given in Table~1, while $\nu_{\rm min} = (3 \, e \, B / 4 \pi \, m_{\rm e} c) \, \gamma_{\rm min}^2 \approx 4.2 \times 10^6 \, (B/{\rm G}) \, \gamma_{\rm min}^2$\,Hz is the synchrotron frequency corresponding to the magnetic field intensity $B$ (measured in Gauss units) and the minimum electron Lorentz factor $\gamma_{\rm min}$. We treat $B$ and $\gamma_{\rm min}$ as free parameters. For such a spectral shape of the synchrotron component \citep[which differs slightly from the ones considered in][]{car91,mei97,wil00,kin04}, we construct the implied electron energy spectrum, $n_{\rm e}(\gamma)$ for $\gamma_{\rm min} < \gamma < \gamma_{\rm max}$, with the normalization $K_{\rm e}$ and the appropriate critical energies $\gamma_{\rm cr}$ and $\gamma_{\rm max}$ depending on $B$ (see \S~4 and equation 3). Next, we compute the SSC emission using the useful approximation formula
\begin{equation}
S_{\nu}^{\rm ssc} \approx {\sqrt{3} \over 4} \, \sigma_{\rm T} \, R \, \nu^{1/2} \, \int_{3 \nu / 4 \gamma_{\rm max}^2}^{3 \nu / 4 \gamma_{\rm min}^2} \, \nu'^{-3/2} \, \, S_{\nu'}^{\rm syn} \, \, n_{\rm e}\!\!\left(\gamma = \sqrt{{3 \, \nu \over 4 \, \nu'}}\right) \, \, d\nu'
\end{equation}
\noindent
\citep[see, e.g.,][]{chi99}, where $R = 0.8$\,kpc is the effective radius of the hotspot, i.e. radius of a sphere with volume equivalent to the cylindrical structure measured directly from the radio maps. All the model parameters and model results discussed in this section are summarized in Table~2.

The evaluated SSC emission at $1$\,keV photon energies is consistent with the observed X-ray flux of hotspot D for $B_{\rm ic} = 270$\,$\mu$G and any minimum electron Lorentz factor $\gamma_{\rm min}$ lower than or equal to $300$. We treat the latter value as the upper limit for $\gamma_{\rm min}$, since Cyg~A hotspots were detected at $<100$\,MHz frequencies \citep{car91,laz06}, while synchrotron radiation of $\gamma = 300$ electrons in $B = 270$\,$\mu$G magnetic field corresponds almost exactly to $\nu = 100$\,MHz. The resulting broad-band synchrotron and SSC emission of hotspot D is shown in Figure~5 (solid lines). We note, that the obtained magnetic field intensity is in the higher end of (although still consistent with) the previous estimates \citep{har94,wil00,kin04,wri04}. This agreement --- despite the slightly different electron spectral shape considered by different authors --- is due to the fact that for the hotspot D parameters, electrons with energies $\gamma \sim 10^4$, i.e. the ones emitting relatively well studied $\sim 100$\,GHz synchrotron emission, are primarily responsible for production of the X-ray SSC flux. It is worth mentioning that our modeling yields also a value of $\alpha^{\rm 0.5\,keV}_{\rm 6\,keV} = 0.93$, which is comparable to the X-ray spectral index given by \citet{wri04}, namely $\alpha_{\rm X} = 0.8\pm0.11$.

Interestingly, with $B_{\rm ic} = 270$\,$\mu$G the expected optical-to-X-ray power-law slope of the SSC emission changes slightly from $\alpha_{\rm O-X} = 0.65$ for $\gamma_{\rm min} = 1$ down to $\alpha_{\rm O-X} = 0.6$ for $\gamma_{\rm min} = 300$. These values are all consistent with the observed one $\alpha_{\rm O-X} \sim 0.6$, and the flux found by \citet[as a residual flux after subtraction of a bright star in the field]{nil97} is overproduced by no more than a factor of $1.8$. This is in fact a crucial result, since the optical SSC emission is produced mostly by the electrons with energies $\gamma < 10^3$, and thus probes the low-energy segment of the electron distribution. Indeed, if one assumes a standard $\nu^{-0.5}$ synchrotron spectrum below the break frequency $\nu_{\rm cr}$ instead of the flat $\propto \nu^{-0.21}$ continuum considered above (arguing that the observed flux at low radio frequencies may be significantly modified by absorption effects), the implied magnetic field intensity needed for producing the X-ray SSC emission would be almost unchanged (namely $B_{\rm ic} = 290$\,$\mu$G), but the optical emission would be overproduced by a factor of $>3$ for $\gamma_{\rm min} = 1 - 300$ (see dotted lines in Figure~5). Thus, the flat synchrotron continuum of hotspot D below $\nu_{\rm cr}$ seems to be intrinsic to the source, and not resulting exclusively from absorption effects. Note in addition, that the choice $\alpha_1 = 0.5$ would lead to overproduction of the $4.5$\,$\mu$m emission as detected by {\it Spitzer}. Of course, inevitable synchrotron self-absorption occurring at $<100$\,MHz frequencies is expected to decrease the predicted SSC flux at low ($<10^{15}$\,Hz) photon energies, and such a decrease should be stronger for steeper synchrotron spectra at low frequencies. However, even in the case of $\alpha_1 = 0.5$ this effect would not remove the aforementioned overproduction of the infrared and optical fluxes of hotspot D.

Finally, we also check if possible steepening of the synchrotron continuum in the `unconstrained' frequency range $>230$\,GHz could somehow change the conclusions presented above. In particular, we repeat the SSC modeling of hotspot D as discussed previously but with an additional spectral break by $\Delta \alpha = 0.5$ at $\nu_{\rm br} = 0.5 \times 10^{12}$\,Hz. In \S~4.2 we argue that around this frequency --- for the particular parameters of hotspot D considered --- one should indeed expect radiative cooling effects leading to steepening of the synchrotron spectrum. The analysis presented in Figure~5 (dashed lines) indicates however that the introduced additional spectral break affects only the unobserved high-energy ($\gg 1$\,keV) part of the expected SSC emission, but not the derived value of $B_{\rm ic}$ nor the low-energy tail of the SSC component.

We follow up the SSC analysis for the case of hotspot A, assuming in a first approach a single-broken spectral shape of its synchrotron continuum. From the collected radio-to-infrared data we infer the input parameters $\alpha_1 = 0.28$, $\alpha_2 = 1.2$, $\nu_{\rm cr} = 2.6$\,GHz, and $R=1.1$\,kpc (see Table~2). Regarding the high-energy cut-off, we find that the value $\nu_{\rm max} = 3.25 \times 10^{13}$\,Hz leads to a rough agreement with the detected $4-8$\,$\mu$m flux, although the implied synchrotron spectrum is then inconsistent with the relatively flat infrared slope constrained by {\it Spitzer}, and it also violates the $3 \sigma$ upper limits by {\it ISO}. In other words, the infrared spectrum of hotspots A cannot be really modelled in terms of high-energy synchrotron cut-off, contrary to the case of hotspot D. Slightly better fit is however obtained by introducing an additional break (as before by about $\Delta \alpha = 0.5$) around $\nu_{\rm br} = 1.2 \times 10^{12}$\,Hz (see \S~4.2), followed by an exponential cut-off at $\nu_{\rm max} \geq 10^{14}$\,Hz. Note, that in such a case the maximum synchrotron frequency cannot be constrained with the available dataset. With the two aforementioned descriptions of the synchrotron spectrum, we find that in both cases the expected SSC radiation matches the observed X-ray flux of hotspot A for $B_{\rm ic} =170$\,$\mu$G and a broad range for $\gamma_{\rm min} = 1 - 300$ (see solid and dashed lines in Figure~6). Clearly, the two different cases considered are energetically equivalent. The evaluated power-law slope $\alpha^{\rm 0.5\,keV}_{\rm 6\,keV} = 1$ is now slightly larger than the appropriate X-ray spectral index provided by \citet{wri04}, $\alpha_{\rm X} = 0.77\pm0.13$. Again, the assumed $\nu^{-0.5}$ synchrotron spectrum below $\nu_{\rm cr}$ leads to slightly larger value of $B_{\rm ic} = 175$\,$\mu$G, but this time to the optical/infrared fluxes well below the available upper limits/detections, and so cannot be formally excluded (see dotted line in Figure~6).

Figure~7 shows the ratio of energy densities in the radiating electrons and magnetic field for hotspots D and A inferred from SSC modeling, $U_{\rm e}/U_{\rm B}$, as a function of the unknown minimum electron Lorentz factor $\gamma_{\rm min}$. One can see that in the case of hotspot D (lower solid line) electrons dominate energetically over the hotspot's magnetic field by only a small factor of $4-3$ for the considered range $\gamma_{\rm min} = 1 - 300$. In the case of hotspot A (upper solid line) this effect is bit stronger, since $U_{\rm e}/U_{\rm B} \sim 8 - 6$ (see also Table~2). Moreover, the $\propto \nu^{-0.5}$ low-frequency synchrotron spectra considered previously for both hotspots as possible (though not likely) case would result in a larger deviation from the minimum energy condition unless $\gamma_{\rm min} > 100$ (lower and upper dotted lines for D and A, respectively). Thus, we conclude that the secondary hotspots in Cyg~A seem to be particle rather than magnetic field dominated \citep[see in this context][]{wil00,wri04,kin04}. However from one perspective the derived energy density ratios are surprisingly close to equality. For example, if some fraction of the observed X-ray emission were to come from some other emission mechanism \citep[e.g., synchrotron; see][]{har04,bal05} then  the calculated value of $B_{\rm ic}$ would become a lower limit on $B$. This would mean fewer electrons to produce the observed radio emission and thus the value of $U_{\rm e}/U_{\rm B}$ would be reduced. Other conditions such as a filling factor $< 1$ would also lead to stronger fields and fewer electrons.

\subsection{Hotspots/Counter-Hotspots Infrared Asymmetry}

The fact that the hotspots A and B on the jet side are much brighter in infrared than hotspots D and E on the counterjet side, while being roughly comparable at other wavelengths (in particular secondary hotspots A and D), raises the question on the nature of the observed asymmetry. There may be several possible reasons for this. One is that due to the intermittent nature of the jet terminal regions and light-travel effects, hotspots on the jet side are observed at a different phase of their evolution than the analogous features on the counterjet side. For the jet viewing angle $\sim 60\deg-70\deg$ in Cyg~A source \citep[see][]{car96}, the jet/hotspot intermittency on timescales $<10^5$\,yrs would be then required to observe any differences in the hotspots' appearance. In fact, \citet{wri04} analyzing the radio data noted that `less spectrally aged material is radiating in hotspot D than in A', consistent with the idea that the observed secondary hotspot on the jet side is older than its counterpart (but see \S~4.2). Yet it is not clear why the more evolved feature should be characterized by higher maximum synchrotron frequency, as suggested by the {\it Spitzer} observations reported here. 

Another possibility is that the time-travel effects/jet intermittency does not play any role, but that instead the electron acceleration process at terminal shocks acts a bit differently on opposite sides of the nucleus. The reason for this could be for example a slightly different configuration of the magnetic filed with respect to the shock front, assuming that the upstream (jet) plasma is moving with at least mildly relativistic velocities. We note, that numerical simulations by \citet{nie04} clearly indicated that the particle spectra (in particular their slopes and maximum energies) resulting from the diffusive (1st order Fermi) acceleration process are indeed very sensitive to the assumed magnetic and kinematic parameters of relativistic shocks. This possibility could be studied by detailed analysis of the Cyg~A hotspot regions at radio frequencies, namely by investigation of the hotspots' polarization properties. We note in this context, that our modeling indicates some differences in \emph{both} magnetic field intensity and high energy synchrotron cut-off between hotspots A and D. This suggests in turn magnetic-related differences in the particle acceleration processes at the highest ($\gamma \sim 10^5$ and above) electron energies.

An interesting issue to note is that the obtained differences in the magnetic field intensity $B_{\rm ic}$ and in the equipartition ratio $U_{\rm e}/U_{\rm B}$ for hotspots A and D do not imply different total kinetic luminosities of the radiating plasma within these two features. Indeed, the total hotspots' energies can be estimated as $E_{\rm tot} = (U_{\rm e} + U_{\rm B}) \times \mathcal{V}$, where $\mathcal{V} = {4 \over 3} \pi \, R^3$. The SSC modeling allows us to find $E_{\rm tot} \sim 1.7 \times 10^{57}$\,erg for hotspot A, and $E_{\rm tot} \sim 0.9 \times 10^{57}$\,erg for its counterpart D (assuming $\gamma_{\rm min} = 1$). Meanwhile, the hotspots' dynamical timescale can be evaluated simply as $t_{\rm dyn} = 2 \, R / \beta c$, where $\beta  c$ is the bulk velocity of the emitting matter. Taking $\beta = 1/3$ as appropriate for the plasma downstream of a strong relativistic shock front, one can find $t_{\rm dyn} \sim 2.2 \times 10^4$\,yrs and $t_{\rm dyn} \sim 1.6 \times 10^4$\,yrs for A and D, respectively. Thus, the implied total kinetic luminosity $L_{\rm kin} = E_{\rm tot} / t_{\rm dyn}$ is very similar for the two analyzed features, being roughly $L_{\rm kin} \sim 2 \times 10^{45}$\,erg s$^{-1}$. We also note that twice this luminosity --- i.e., total power carried by the radiating electrons and magnetic field in both hotspots --- is only slightly lower than total kinetic power inferred from expansion of the Cyg~A cavity in the surrounding cluster gas, $\sim 10^{46}$\,erg s$^{-1}$ \citep{wil06}.

\section{Constraining The Electron Energy Distribution}

Electron spectra of hotspots D and A within the energy range $300 < \gamma < 30000$ (not affected by absorption or cooling effects), as found from the SSC modeling presented above, can be approximated by a broken power-law
\begin{equation}
n_{\rm e} (\gamma) = K_{\rm e} \times \left\{ \begin{array}{ccc} \gamma^{- p_1} & {\rm for} &
 \gamma < \gamma_{\rm cr} \\ \gamma_{\rm cr}^{p_2-p_1} \, \gamma^{- p_2} &
 {\rm for} & \gamma > \gamma_{\rm cr} \end{array} \right. \quad ,
\end{equation}
\noindent
with $\gamma_{\rm cr} \approx 2 \times 10^3$, $p_1 = 2 \, \alpha_1 +1 \approx 1.4-1.6$, and $p_2 = 2 \, \alpha_2 +1 \approx 3.2-3.4$. At higher energies the electron distributions --- possibly modified by the radiative loss effect --- seem to continue up to at least $\gamma \approx 10^5$. These spectra are shown in Figure~8 (solid and dashed lines correspond to hotspots D and A, respectively; see also Table~2). We emphasize that
\renewcommand\theenumi{(\roman{enumi})} 
\begin{enumerate}
\item for both hotspots the critical break energy corresponds almost exactly to the mass ratio between protons and electrons, $\gamma_{\rm cr} \approx m_{\rm p}/m_{\rm e}$, which gives a natural scale unit in the case of a shock front formed in an electron-proton plasma; 
\item for both hotspots the low-energy segment of the electron distribution continues down to at least $\gamma_{\rm min} \sim 0.1 \, m_{\rm p}/m_{\rm e}$ electron energies, with flat power-law spectrum of slope $p_1 \approx 1.5$ possibly slightly different in the two analyzed cases;
\item for both hotspots the high-energy part of the electron continuum is very steep, with spectral indices $p_2 > 3$ and maximum energies $\gamma_{\rm max} \gtrsim 50 \, m_{\rm p}/m_{\rm e}$ slightly different for the two features;
\item in both hotspots `standard' electron spectrum $n_{\rm e}(\gamma) \propto \gamma^{-2}$ expected from the diffusive (1st order Fermi) shock acceleration in the non-relativistic test-particle limit is not observed.
\end{enumerate}
The latter finding regarding the model shock-type spectrum $\propto \gamma^{-2}$ is an interesting issue. It should not be surprising though, keeping in mind the fact that the terminal shocks of powerful jets, like those of the Cyg~A radio galaxy analyzed here, are almost for certain (mildly) relativistic, with an oblique magnetic field configuration. As such, they are indeed not expected to be the sites of the diffusive acceleration process known from the non-relativistic test-particle models \citep[see the relevant discussion in][]{beg90,nie06,lem06}. The break frequency corresponding to the $m_{\rm p}/m_{\rm e}$ mass ratio preceded by a flat-spectrum power-law is also an interesting but again not a very unique \citep[see][]{lea89} nor completely unexpected finding. As discussed below, it may imply several important constraints on extragalactic jets in general.

\subsection{Low-Energy Spectrum}

The question of accelerating the electrons from thermal (non- or mildly-relativistic) to ultrarelativistic energies at relativistic shocks is widely debated. That is because the electrons have to be already ultrarelativistic to undergo diffusive 1st-order acceleration, since otherwise their gyroradii are smaller than the thickness of the velocity transition region, and thus the low-energy particles cannot cross the shock front freely. This issue, known as the `injection problem' \citep[e.g.,][]{bel78a,bel78b,eil90}, is of particular relevance for the collisionless plasma which is dynamically dominated by protons, because the shock thickness is then expected to be of the order of the dominant protons' gyroradius. For the upstream matter being cold in its rest frame, the required `injection' electron Lorentz factors would be roughly $\sim m_{\rm p}/m_{\rm e}$. It was postulated, that some efficient but unspecified energy/temperature coupling acting between protons and electrons (which cannot be collisional in nature) should serve as the pre-acceleration mechanism. Yet the low-energy ($\gamma < m_{\rm p}/m_{\rm e}$) electron spectrum resulting from such a process remained unknown. The Cyg~A hotspots are, obviously, subject to this problem \citep[see the discussion in][]{car91,kin04}.

In their pioneering work, \citet{hos92} presented a possible solution to the injection problem for the case of a cold electron-proton plasma upstream of the shock. In particular, by using 1D particle-in-cell simulations, they showed that within the velocity transition region positrons can absorb electromagnetic waves emitted at high harmonics of the cyclotron frequency by the cold protons reflected from the shock front, and thus be accelerated to suprathermal energies. For efficient acceleration it was required to assume at least comparable energy densities in protons and in electron-positron pairs. This result concerned strictly positrons and not electrons, because for the parameters considered by \citeauthor{hos92} polarization of the proton cyclotron waves was almost exclusively left-handed. Recently, however, \citet{ama06} --- who applied more realistic parameters in their numerical studies when compared with simulations by \citet{hos92} --- showed that the same process may as well apply to the electrons\footnote{For the parameters used in the 1D particle-in-cell simulations reported by \citet{hos92} --- small proton-to-electron mass ratio $m_{\rm p}/m_{\rm e} \approx 10$ and similar energy densities in leptons and hadrons --- the number of protons was comparable to the number of pairs, and thus polarization of the protons' cyclotron emission was almost completely left-handed. \citet{ama06} considered much larger proton-to-electron mass ratio $m_{\rm p}/m_{\rm e} \approx 100$ which, for comparable energy densities in the two plasma species, corresponds to a much smaller number of protons than pairs. Under such conditions, proton cyclotron emission is expected to be a mixture of left- and right-handed modes, which then can be absorbed by both electrons and positrons.}. The resulting electrons spectra are consistent with a power-law between energies $\gamma \sim \Gamma$, where $\Gamma$ is the bulk Lorentz factor of the upstream medium, and $\gamma \sim \Gamma \, (m_{\rm p}/m_{\rm e})$, possibly reduced by a factor of a few due to thermal dispersion in the upstream proton momenta. The spectral index of this distribution, hardly constrained by the simulations, seems to depend on the plasma content (i.e., on the ratio of proton number density to the electron number density), and can be relatively flat. We emphasize that although simulations presented by \citet{hos92} and \citet{ama06} involve large bulk Lorentz factors of the upstream plasma, analogous processes are expected to take place also at mildly relativistic shocks considered in this paper ($\Gamma \sim \Gamma_{\rm j}\sim few$), so the model results can be qualitatively applied.

Our analysis indicates that a flat power-law electron spectrum $\propto \gamma^{-1.5}$ in Cyg~A hotspots extends from low (unfortunately unconstrained) electron energies up to almost exactly $\gamma_{\rm cr} \approx m_{\rm p}/m_{\rm e}$. This is in fact in a good agreement with the simple (1D, ultrarelativistic) resonant acceleration model discussed by \citet{ama06}. If such an association is correct, and if the results of \citeauthor{ama06} as applicable, then it would automatically imply that {\it (i)} the jets in powerful radio sources like Cyg~A are made of electron-proton rather than electron-positron plasma, {\it (ii)} a significant fraction of jet kinetic power is carried by cold protons, and {\it (iii)} number density of protons within the jets is most likely lower than that of the electron-positron pairs. We note, that the above conclusions regarding the jet content would be then consistent with the ones presented by \citet{sik00}, who analyzed the broad-band spectra and variability of radio-loud quasars, believed to be beamed counterparts of FR~II radio galaxies \citep[see also][who presented other evidence for the dynamical role of protons in quasar jets]{ghi02}. 

\subsection{High-Energy Spectrum}

It was stated in previous studies \citep{car91,mei97}, that radiative cooling effects are of importance for the electrons emitting $\gtrsim 1-10$\,GHz synchrotron photons within the Cyg~A hotspots A and D, and hence that the spectrum of `freshly' accelerated electrons in the appropriate energy range is flatter by some $\Delta p$ than the observed (cooled) one. In a framework of the `continuous injection' model \citep[widely applied to hotspots in general in spite of the obvious fact that the CI model rests on the premise that particles accumulate in the emitting volume whereas hotspots are emitting volumes characterized by both a hefty source function and a strong outflow; see][]{hea87,mei89}, such a corrected electron spectral index would be then $\bar{p}_2 = 2.2-2.4$ instead of the observed one $p_2 = 3.2-3.4$, assuming a homogeneous distribution of the magnetic field within the emitting volume leading to $\Delta \alpha = 0.5$. If this is the case, then fine-tuning of the model parameters would be required in order to equalize the observed critical energy $\gamma_{\rm cr} \sim m_{\rm p}/m_{\rm e}$, argued above to be intrinsic to the electron spectrum, with the break energy resulting from cooling effects, denoted here as $\gamma_{\rm br}$. This, as discussed below, is however not expected.

The synchrotron break frequency corresponding to the transition between slow- and fast-cooling regime, $\nu_{\rm br} = (3 \, e \, B / 4 \pi \, m_{\rm e} c) \, \gamma_{\rm br}^2$, can be found by equalizing the dynamical timescale introduced previously, $t_{\rm dyn} = 2 \, R / \beta \, c$, with the energy-dependent radiative cooling timescale $t_{\rm rad}(\gamma)$. The latter one can be found as $t_{\rm rad} = \gamma / |\dot{\gamma}|_{\rm syn}$, where the (dominant) synchrotron cooling rate is simply $|\dot{\gamma}|_{\rm syn} = c \sigma_{\rm T} \, B^2 \, \gamma^2 / (6 \pi \, m_{\rm e} c^2)$. Hence, the condition $t_{\rm dyn} = t_{\rm rad}(\gamma_{\rm br})$ gives
\begin{equation}
\nu_{\rm br} \approx {27 \pi \, m_{\rm e} c^3 \, e \, \beta^2 \over 4 \, \sigma_{\rm T}^2 \, B^3 \, R^2} \approx 6 \times 10^{13} \, \beta^2 \, B_{-4}^{-3} \, R_{\rm kpc}^{-2} \quad {\rm Hz} \quad ,
\end{equation}
\noindent
where $B_{-4} \equiv B/10^{-4}$\,G, and $R_{\rm kpc} \equiv R/1$\,kpc \citep[see][]{bru03,che05}. Taking hotspots' parameters as discussed in previous sections, one obtains $\nu_{\rm br} =0.5 \times 10^{12}$\,Hz for hotspot D and $\nu_{\rm br} =1.2 \times 10^{12}$\,Hz for hotspot A. This is indeed much higher than the observed critical frequency $\nu_{\rm cr} \approx 3 \times 10^9$\,Hz. Hence we conclude that the observed steep electron spectrum at $\gamma \geq m_{\rm p}/m_{\rm e}$ results directly from the acceleration process, and is not affected by the radiative cooling effects. We note that \citet{kin04} evaluated the cooling break frequency as high as discussed above. The discrepancy with the previous works, like that by \citet{car91} and \citet{mei97} is mainly due to a much lower outflow velocity assumed by these authors for the emitting region downstream of the terminal shock front. This resulted in a larger (by more than an order of magnitude) dynamical timescale for the hotspot, and thus a lower cooling break frequency. From equation 4 it follows directly that the outflow velocity should be reduced as much as $\beta < 0.03$ in order to obtain $\nu_{\rm br} < 10$\,GHz. Even keeping in mind the expected very complex structure of the outflowing plasma in the discussed region, such low downstream velocities seem to be not realistic. Note, that typical advance velocities of hotspot features in FR~II radio galaxies --- which have to be lower than $\beta$ discussed here --- are typically $\beta_{\rm adv} \gtrsim 0.01$ \citep[e.g.,][]{mac07}.

If we consider exclusively the case where diffusive shock acceleration is responsible for the formation of the high energy electron spectra observed in the Cyg~A hotspots, we have to accept a significant departure of the electron spectral index from the value $p_2 = 2$ corresponding to the non-relativistic shock, but also from the often claimed `universal' value $p_2 = 2.2$ corresponding to the (ultra)relativistic shock. In fact, recent modeling of the first-order Fermi acceleration process at relativistic shocks \citep{nie04,nie06} indicates the possibility for formation of extended high energy `power-law +  cut-off' tails above the shock-related $\Gamma \, m_{\rm p} c^2$ energy scale. The theory is however not able to provide robust evaluations of the involved spectral indices and maximum energies yet, and thus the analysis presented here could be considered as valuable empirical measurements of these parameters for mildly relativistic shocks. We emphasize that the observed differences in the power-law slopes and cut-off frequencies between the two hotspots may then indicate sensitivity of the particle spectra resulting from the 1st order Fermi acceleration process on (even relatively small) differences in the magnetic field configuration and intensity within the shock front. We note in this context, that hotspot A shows a larger deviation from the minimum power condition than hotspot D.

The second possibility we consider is that of the shock responsible for accelerating the electrons only below $\sim \Gamma \, m_{\rm p} c^2$ energies (as discussed in \S~4.1), and of a distributed acceleration in the turbulent hotspot volume responsible for accelerating electrons above that scale. The acceleration processes involved can be the second-order Fermi acceleration by relativistic hotspot turbulence downstream of the jet terminal shock, or particle acceleration accompanying magnetic field reconnection events. Both these processes were hardly studied for the particular hotspots' conditions, but the appropriate rough time scale evaluations easily allow for electron energies as high as the $\gamma_{\rm max}$ values required here. As illustrative examples of recent theoretical progress in this field one can compare \citet{pet04,cho06} for the second-order acceleration, and \citet{jar04,lyu05,zen05} for the reconnection process. Also in this case the electron spectral slope derived from the observations could serve as an important input for better understanding of the aforementioned acceleration processes.

\section{Conclusions}

We have presented the results of {\it Spitzer Space Telescope} observations of the famous radio galaxy Cygnus~A. We have detected both hotspots A and B on the west (jetted) side at $4.5$\,$\mu$m and $8$\,$\mu$m. Because confusing emission precludes accurate estimates of the background level, our measurements of D and E on the eastern side are of low statistical significance. In all cases the observed infrared emission is consistent with the high-energy tail of the hotspots' synchrotron radiation. When combined with the other data collected from the literature, our observations allow for detailed modeling of the broad-band (radio-to-X-ray) emission for the brightest hotspots A and D, uniquely sampled at different wavelengths. We confirm that the X-ray emission detected previously by {\it ROSAT} and {\it Chandra} satellites is consistent with the synchrotron self-Compton radiation for the magnetic field intensity $B \approx 170$\,$\mu$G in hotspot A and $B \approx 270$\,$\mu$G in hotspot D. However, we also find that the energy density of the emitting electrons is larger than the energy density of the hotspots' magnetic field by a factor of a few, and that hotspot A is more particle dominated than hotspot D. Finally, we note that the hotspots located on the jet side (A and B) are significantly brighter in the infrared than their counterparts located on the other side of the active center (D and E). We suggest that such an asymmetry is most likely due to different configurations and different relative strengths of the magnetic field within the terminal shocks on the jet and counterjet sides. These, in turn may result either from temporal evolution of the hotspots (coupled with the fact we are observing them at different proper ages), from differences in the cluster medium on the two sides, or from differences in the jet energy or field configuration on timescales of $10^5$\,yrs.

The performed analysis allows us to constrain relatively precisely the energy spectra of ultrarelativistic electrons producing the broad-band emission in hotspots A and D. We find that in both cases it is consistent with a flat power-law $n_{\rm e}(\gamma) \propto \gamma^{-p_1}$ with $p_1 = 1.4-1.6$ for the electron energies $\gamma < m_{\rm p}/m_{\rm e}$, followed by a steep power-law $n_{\rm e}(\gamma) \propto \gamma^{-p_2}$ with $p_2 = 3.2-3.4$ for $\gamma > m_{\rm p}/m_{\rm e}$. We argue that such a shape of the electron continuum reflects most likely two different regimes of the electron acceleration process taking place at mildly relativistic shock, rather than resulting from the radiative cooling and/or absorption effects. In this picture the protons' inertia defines a critical energy for the hotspot electrons above which Fermi-type acceleration processes may play a major role, but below which the operating acceleration mechanism has to be of a different type. We suggest that the latter mechanism is connected with collisionless processes acting within the shock transition layer, most likely involving resonant absorption of the shocked ions' cyclotron waves by the electron-positron pairs \citep{hos92,ama06}. If this interpretation is correct, it would automatically imply a dynamical role of (cold) protons within the jet, consistent with analyses of jet energetics in quasar sources. It is important to note that the hotspots in other powerful radio sources analogous to Cyg~A reveal similar spectral breaks around GHz frequencies \citep[see][]{lea89}. Obviously, insufficient understading of the relativistic shock structure, as well as limitations of the available numerical methods in modeling microphysics of relativistic magnetized plasma, leave some room for the other processes which can be responsible for (pre)acceleration of low-energy electrons. Let us mention in this context the interaction of particles with electrostatic waves excited at the front of quasi-perpendicular shocks \citep{sch00,hos02}, or with quasi-stationary electric field associated with the two-stream shock instabilities \citep{hed04,nis06}. We note that any type of turbulence inevitably generated in e$^-$p$^+$ shock by protons will lead to acceleration of electrons up to the downstream proton energies, but only a few simple models regarding this type of processes were studied till now.

The high-energy part of the electron distribution above $m_{\rm p} / m_{\rm e}$ energies may result from the Fermi-type processes, although the steep slope of the electron spectrum at these energies deviates significantly from the `universal' spectral indices often claimed in the literature for diffusive shock acceleration. We argue however that such a disagreement is in fact only superficial, since the most recent theoretical studies on the particle acceleration taking place at relativistic and oblique shocks clearly indicate an absence of any universal spectrum, but instead show a variety of particle spectral shapes. This is supported further by the observed differences in the electron spectral shape between hotspots A and D at the highest electron energies, $> 50 \, m_{\rm p} / m_{\rm e}$, which indicates that the 1st order Fermi acceleration process --- if indeed responsible for the formation of the high-energy electron tail --- is very sensitive to (even relatively) small differences in the magnetic field structure at the shock front. We also note that our conclusions are consistent with the general finding that the electron spectrum injected from the terminal hotspots to the lobes of powerful FR~II radio galaxies is not of a single and universal power-law form \citep[as discussed in detail by, e.g.,][]{rud94,tre01,mac07}. This, often ignored result, has several consequences for estimating the jet lifetimes and thus energetics and duty-cycles of radio-loud active galactic nuclei.

\acknowledgments

\L .S. and M.O. were supported by MEiN through the research project 1-P03D-003-29 in years 2005-2008. \L .S. acknowledges K. Katarzy\'nski and M. Kino for helpful comments. D.E.H. acknowledges useful discussions with J. Lazio and J. Eilek. This work is based on observations made with the {\it Spitzer Space Telescope}, which is operated by the Jet Propulsion Laboratory, California Institute of Technology under a contract with NASA. The work at SAO was partially supported by Spitzer grant 1279229.  The National Radio Astronomy Observatory is operated by Associated Universities, Inc. under a cooperative agreement with the National Science Foundation. The authors thank the anonymous referee for helpful suggestions.

{}

\begin{deluxetable}{llllll}
\tabletypesize{\scriptsize}
\tablecaption{Spectral flux densities for the Cyg~A hotspots.}
\tablewidth{0pt}
\tablehead{
\colhead{} 
& \colhead{Hotspot A}
& \colhead{Hotspot B}
& \colhead{Hotspot D}
& \colhead{Hotspot E}
& \colhead{}\\
$\nu$ [Hz] & $S_{\nu}$ [Jy] & $S_{\nu}$ [Jy] & $S_{\nu}$ [Jy] & $S_{\nu}$ [Jy] & ref.}
\startdata
0.327 E9 & 173$\pm$25 & 50$\pm$15 & 162$\pm$20 & --- & [1]\\
1.34 E9 & 116.4$\pm$5.8 & 31.5$\pm$3.0 & 120.6$\pm$6.0 & 10$\pm$1.0 & [2]\\
5.0 E9 & 44.60$\pm$4.46 & 9.96$\pm$1.00 & 57.69$\pm$5.77 & 3.90$\pm$0.39 & [3]\\
15 E9 & 10.80$\pm$1.08 & 2.47$\pm$0.25 & 17.74$\pm$1.77 & 1.16$\pm$0.12 & [3]\\
87 E9 & 1.49$\pm$0.18 & 0.44$\pm$0.05 & 2.49$\pm$0.30 & 0.19$\pm$0.02 & [3]\\
230 E9 & 0.43$\pm$0.05 & 0.11$\pm$0.01 & 0.87$\pm$0.10 & 0.10$\pm$0.01 & [3]\\
2.5 E13 & $<$ 285 E-6 & $<$ 285 E-6 & $<$ 255 E-6 & $<$ 255 E-6 & [4]\\
3.798 E13 & 156$\pm$19 E-6 & 50$\pm$12 E-6 & 31$\pm$25 E-6 & 15$\pm$24 E-6 & [5]\\
6.655 E13 & 75$\pm$9 E-6 & 25.3$\pm$18 E-6 & 5.5$\pm$5.5 E-6 & 3.5$\pm$7 E-6 & [5]\\
1.36 E14 & $<$  30$\pm$3 E-6 & $<$ 35$\pm$2 E-6 & $<$ 25 E-6 & --- & [6]\\
2.48 E14 & $<$ 100$\pm$10 E-6 & $<$ 52.4$\pm$2.6 E-6 & --- & --- & [6]\\
4.29 E14 & --- & --- & 1.63$\pm$0.3 E-6 & --- & [7]\\
4.63 E14 & $<$ 80$\pm$8 E-6 & $<$ 46.2$\pm$0.3 E-6 & --- & --- & [6]\\
5.45 E14 & --- & 2.65$\pm$0.6 E-7 & --- & --- & [7]\\
6.64 E14 & $<$ 50$\pm$5 E-6 & $<$ 21.9$\pm$0.8 E-6 & $<$ 5 E-6 & --- & [6]\\
2.42 E17 & 31.2$\pm$4.3 E-9 & 6.8$\pm$2.6 E-9 & 47.9$\pm$5.9 E-9 & --- & [8]\\
\hline
$\alpha_{\rm X}$ & 0.77$\pm$0.13 & 0.70$\pm$0.35 & 0.80$\pm$0.11 & --- & [8]\\
 & (0.5$-$7.5\,keV) & (0.5$-$4.5\,keV) & (0.5$-$6.0\,keV) &  & 
\enddata
\tablecomments{ 
References: [1] T.J.W. Lazio, private communication \citep[see also][]{laz06}; [2] \citet{car91}; [3] \citet{wri04}; [4] \citet[$3 \sigma$ upper limits]{car99}; [5] this work; [6] \citet{mei97}; [7] \citet{nil97}; [8] \citet[corrected for an additional intrinsic absorption]{wri04}}
\end{deluxetable}

\clearpage

\begin{deluxetable}{lcccc}
\tablecaption{Model Parameters for the Cyg~A hotspots}
\tablewidth{0pt}
\tablehead{
\colhead{}& \multicolumn{2}{|c}{Hotspot A}	&\multicolumn{2}{|c}{Hotspot D}} 
\startdata
\\
\multicolumn{5}{c}{Measured Parameters}\\
\hline
$\alpha_{\rm 1}$        &\multicolumn{2}{|c}{0.28}           		&\multicolumn{2}{|c}{0.21}   \\
$\alpha_{\rm 2}$        &\multicolumn{2}{|c}{1.2}            		&\multicolumn{2}{|c}{1.1}    \\
$\nu_{\rm cr}$ [Hz]     &\multicolumn{2}{|c}{2.6$\times$10$^{9}$}   	&\multicolumn{2}{|c}{3.1$\times$10$^{9}$}  \\
$\nu_{\rm max}$ [Hz]    &\multicolumn{2}{|c}{$>$3.3$\times$10$^{13}$} 	&\multicolumn{2}{|c}{0.9$\times$10$^{13}$} \\ 
Cylindrical dimensions ($L,r$)	&\multicolumn{2}{|c}{(2.2'', 0.8'')}	&\multicolumn{2}{|c}{(1.9'', 0.7'')}\\
Equivalent radius $R$ [kpc]&\multicolumn{2}{|c}{1.1}   		&\multicolumn{2}{|c}{0.8}    \\ 
\hline
\\ 
\multicolumn{5}{c}{Fitted Parameters}\\
\hline
$\gamma_{\rm min}$&\multicolumn{1}{|c}{1}&\multicolumn{1}{|c}{300}&\multicolumn{1}{|c}{1}
&\multicolumn{1}{|c}{300}\\
$\alpha_{\rm X}$        &\multicolumn{1}{|c}{1}              &\multicolumn{1}{|c}{1}       
&\multicolumn{1}{|c}{0.93}           &\multicolumn{1}{|c}{0.93}\\
$\alpha_{\rm O-X}$      &\multicolumn{1}{|c}{--} &\multicolumn{1}{|c}{--}        
&\multicolumn{1}{|c}{0.65}           &\multicolumn{1}{|c}{0.60}\\
$\gamma_{\rm cr}$       &\multicolumn{1}{|c}{1.9$\times$10$^{3}$} &\multicolumn{1}{|c}{1.9$\times$10$^{3}$} 
&\multicolumn{1}{|c}{1.7$\times$10$^{3}$}  & \multicolumn{1}{|c}{1.7$\times$10$^{3}$}\\ 
$\gamma_{\rm max}$      &\multicolumn{1}{|c}{$>$2.2$\times$10$^{5}$}& \multicolumn{1}{|c}{$>$2.2$\times$10$^{5}$}       
&  \multicolumn{1}{|c}{0.9$\times$10$^{5}$}             &\multicolumn{1}{|c}{0.9$\times$10$^{5}$} \\
$K_{\rm e}$ [cm$^{-3}$]  &\multicolumn{1}{|c}{1.4$\times$10$^{-4}$} & \multicolumn{1}{|c}{1.4$\times$10$^{-4}$}  &
\multicolumn{1}{|c}{7.6$\times$10$^{-5}$} &\multicolumn{1}{|c}{7.6$\times$10$^{-5}$}\\ 
$B_{\rm ic}$ [$\mu$G]           &\multicolumn{1}{|c}{170}&\multicolumn{1}{|c}{170}&\multicolumn{1}{|c}{270}
&\multicolumn{1}{|c}{270}\\
$U_{\rm e}/U_{\rm B}$   &\multicolumn{1}{|c}{8}               & \multicolumn{1}{|c}{7}       &               
\multicolumn{1}{|c}{4} &\multicolumn{1}{|c}{3}\\
\enddata
\tablecomments{The spectral parameters, $\alpha_{\rm 1}$, $\alpha_{\rm 2}$, $\nu_{\rm cr}$, and $\nu_{\rm max}$, are defined in equation~1. We estimated the length $L$ and radius $r$ of a cylindrical volume for the hotspots; for convenience, we give the equivalent radii $R$ of spheres of equal volume to the cylinders.\\
From the SSC modeling, we infer the magnetic field $B$ and report two spectral indices of the SSC spectrum which can be compared with observations ($\alpha_{\rm O-X}$, and $\alpha_{\rm X}$ which is between 0.5 and 6 keV). We constrain also the parameters of the electron energy spectrum ($\gamma_{\rm cr}$, $\gamma_{\rm max}$, and $K_{\rm e}$; see equation~3), as well as the ratios of the energy densities $U_{\rm e}/U_{\rm B}$.}
\end{deluxetable}

\clearpage

\begin{figure}
\includegraphics[scale=0.55,angle=0]{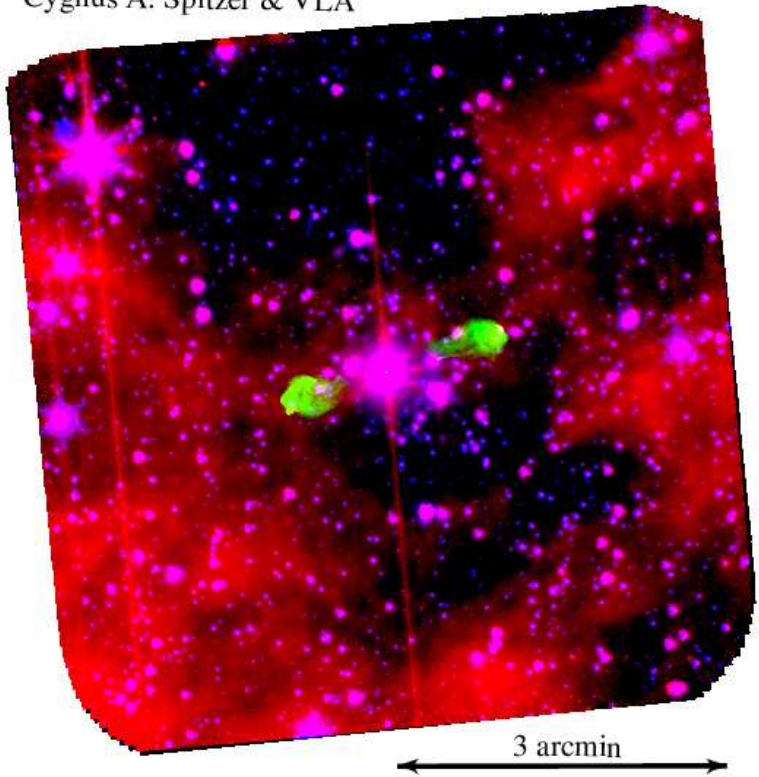}
\caption{{\it Spitzer Space Telescope} IRAC and VLA image \citep[green; from][]{per84} of Cygnus~A. The 
4.5$\mu$m sources are shown in blue, 8.0$\mu$m emission in red and those emitting in 
both bands in pink.} \end{figure}

\begin{figure}
\includegraphics[scale=0.45,angle=270]{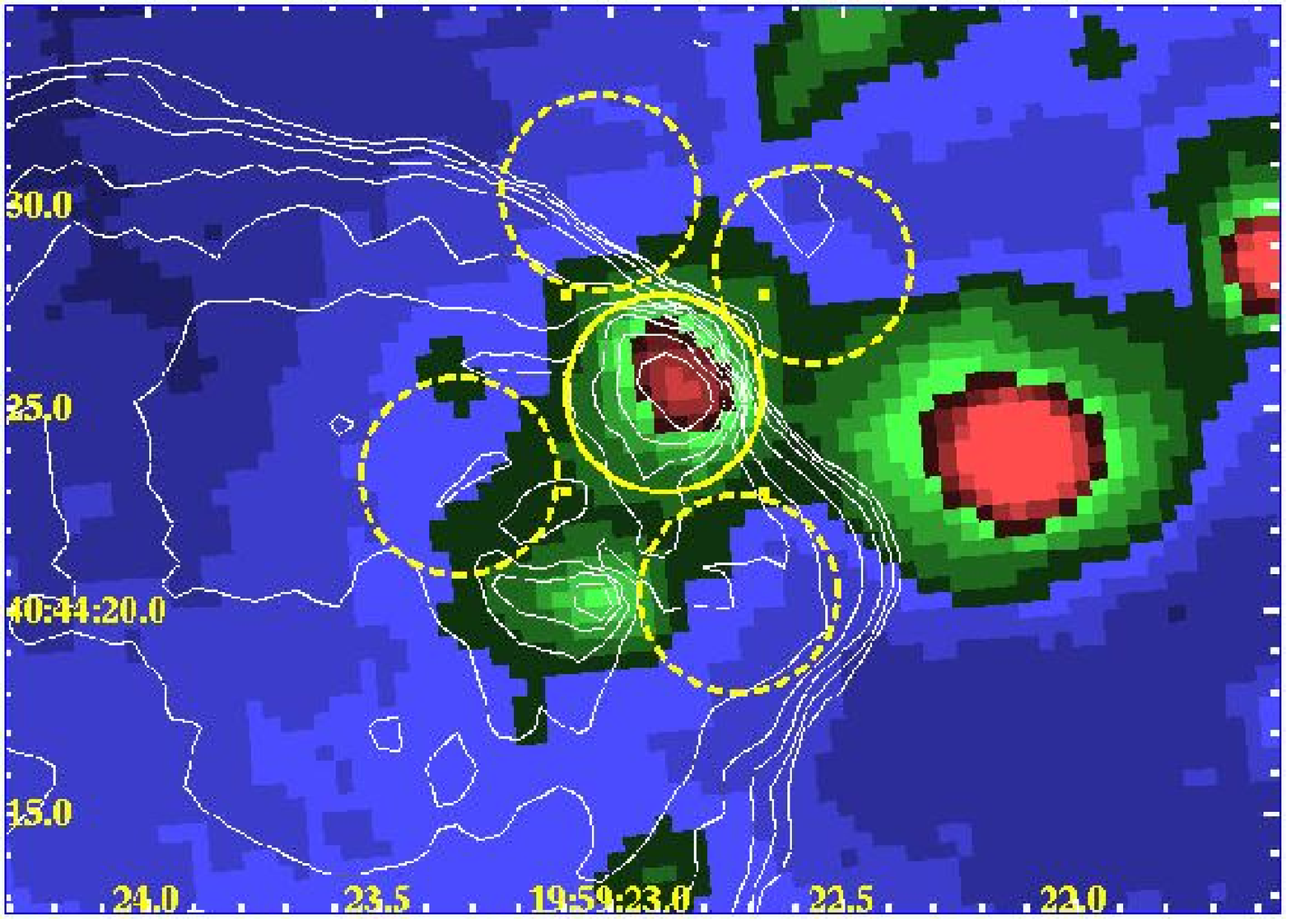}\\
\includegraphics[scale=0.45,angle=270]{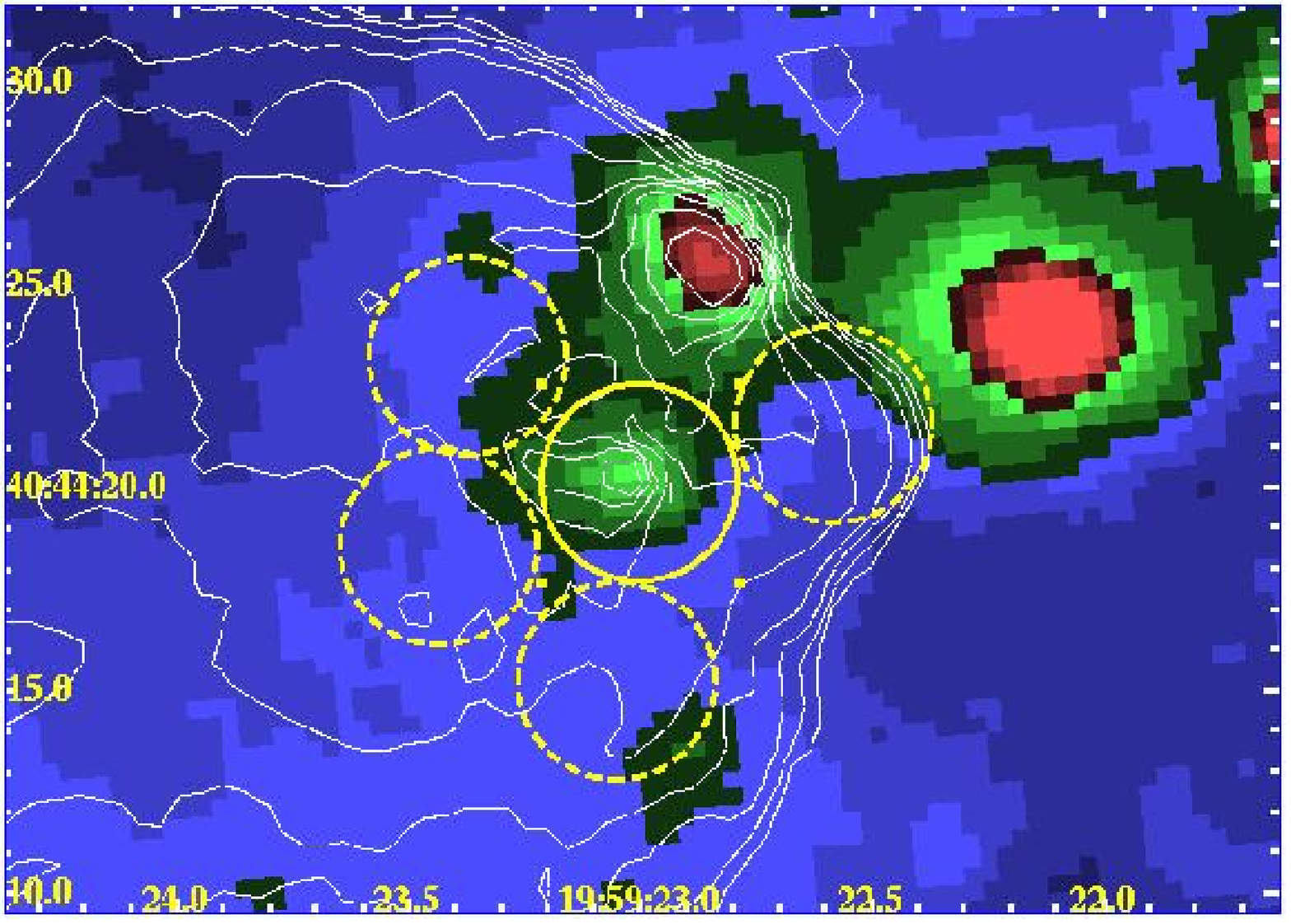}\\
\includegraphics[scale=0.45,angle=270]{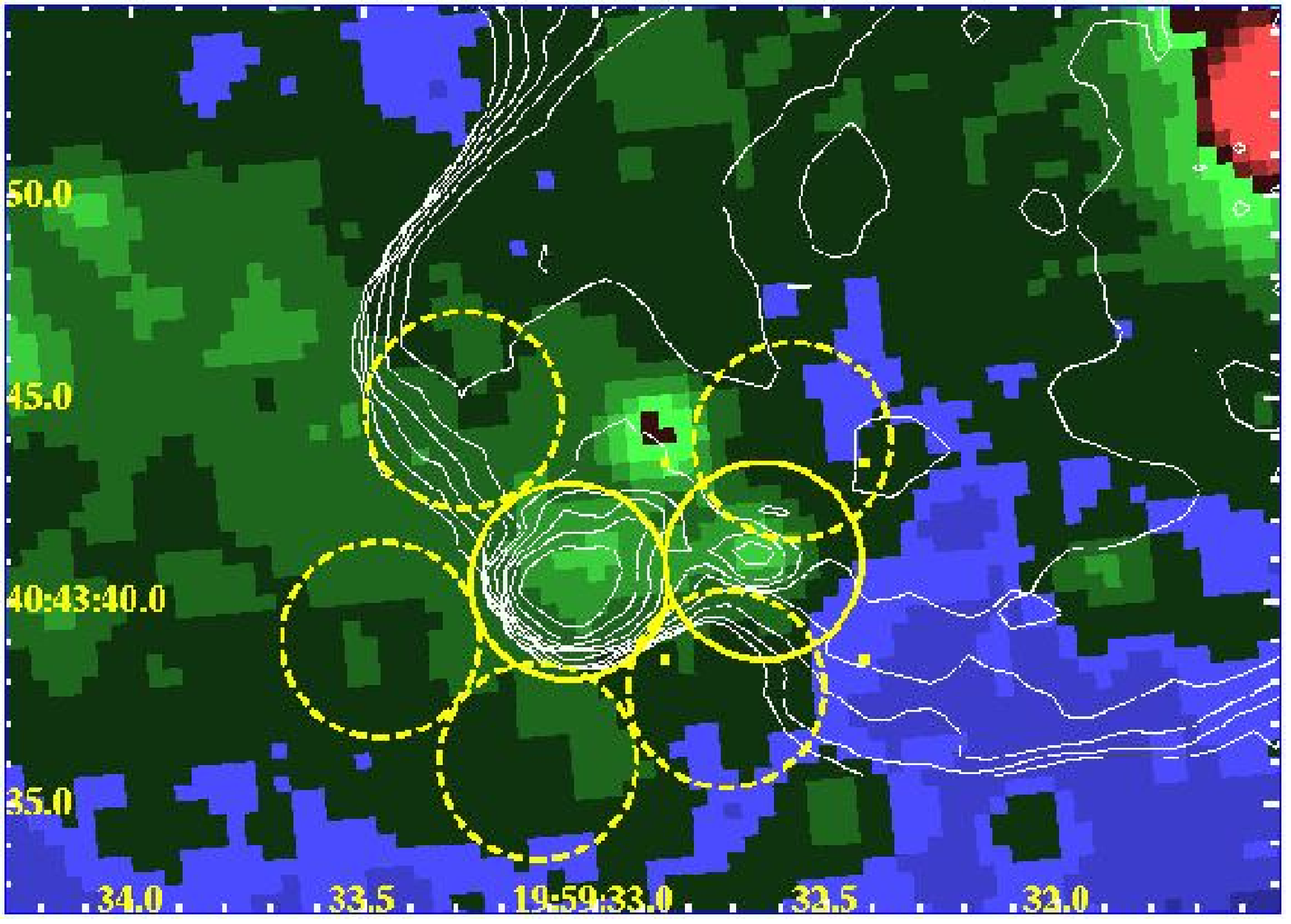}
\caption{$8$\,$\mu$m images of Cyg~A hotspots with $5$\,GHz contours superimposed. Regions used for the hotspots' and backgrounds' flux subtraction are marked by solid and dashed circles, respectively. {\it Upper:} Secondary hotspot A on the jet side. {\it Middle:} Primary hotspot B on the jet side. {\it Lower:} Secondary hotspot D (left) and primary hotspot E (right) on the counterjet side.}
\end{figure}

\begin{figure}
\includegraphics[scale=1.50]{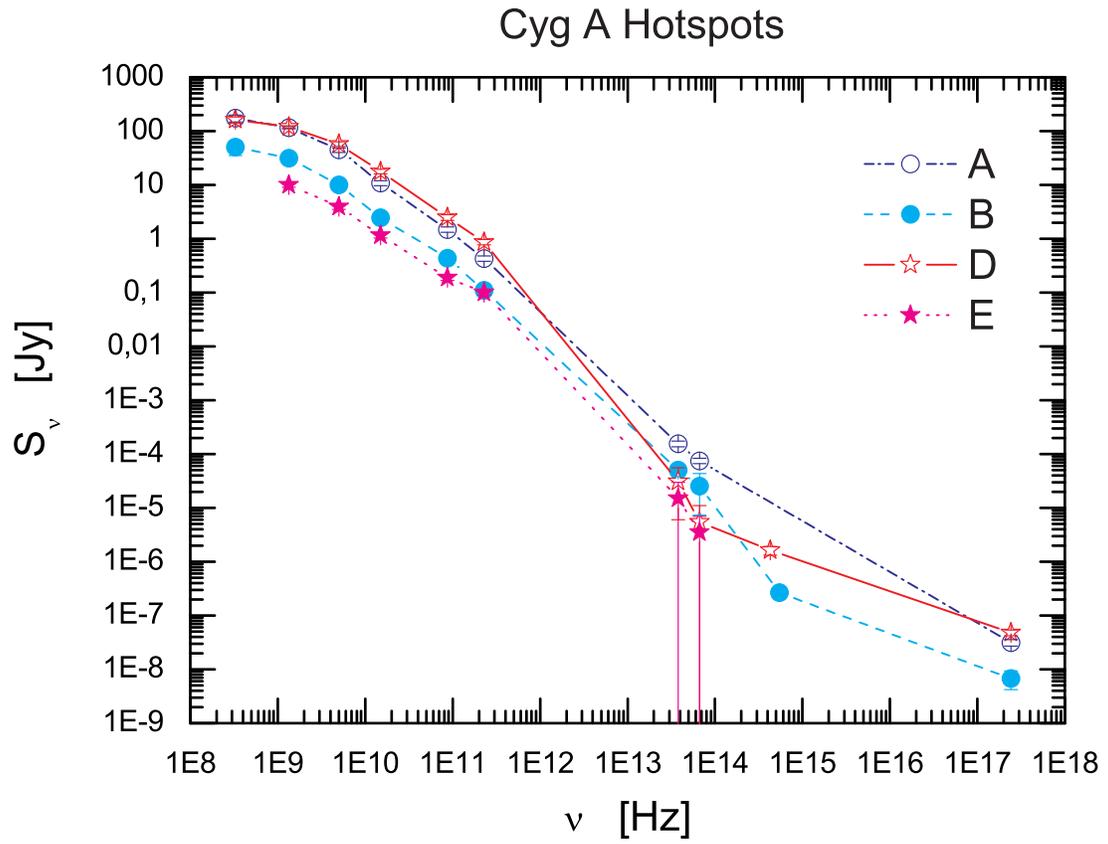}
\caption{Flux densities for Cyg~A hotspots (see Table~1). Lines do not correspond to any model fit, but only connect data points for each hotspot.}
\end{figure}

\begin{figure}
\includegraphics[scale=1.50]{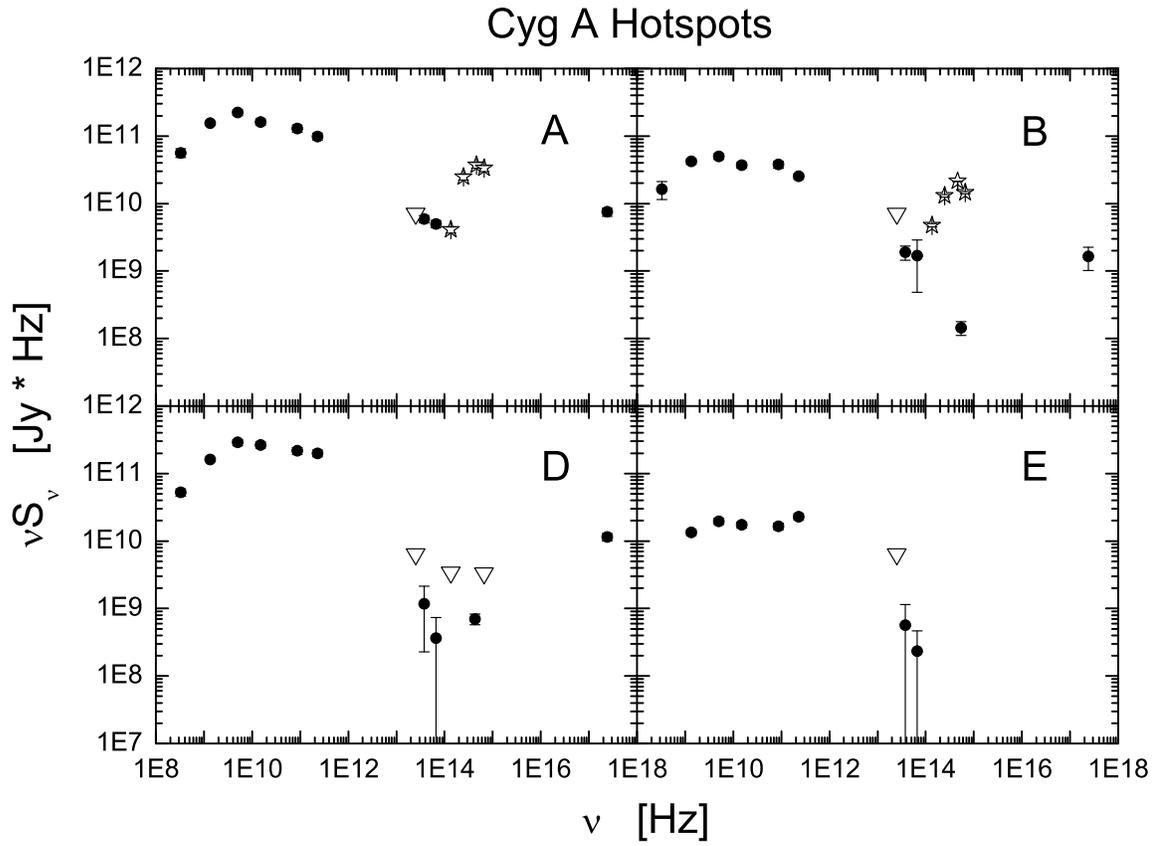}
\caption{Spectral energy distribution of the broad-band emission observed from Cyg~A hotspots. Open triangles denote upper limits, open stars correspond to known foreground star in the field, while filled circles correspond to the detected fluxes (see Table~1).}
\end{figure}

\begin{figure}
\includegraphics[scale=1.50]{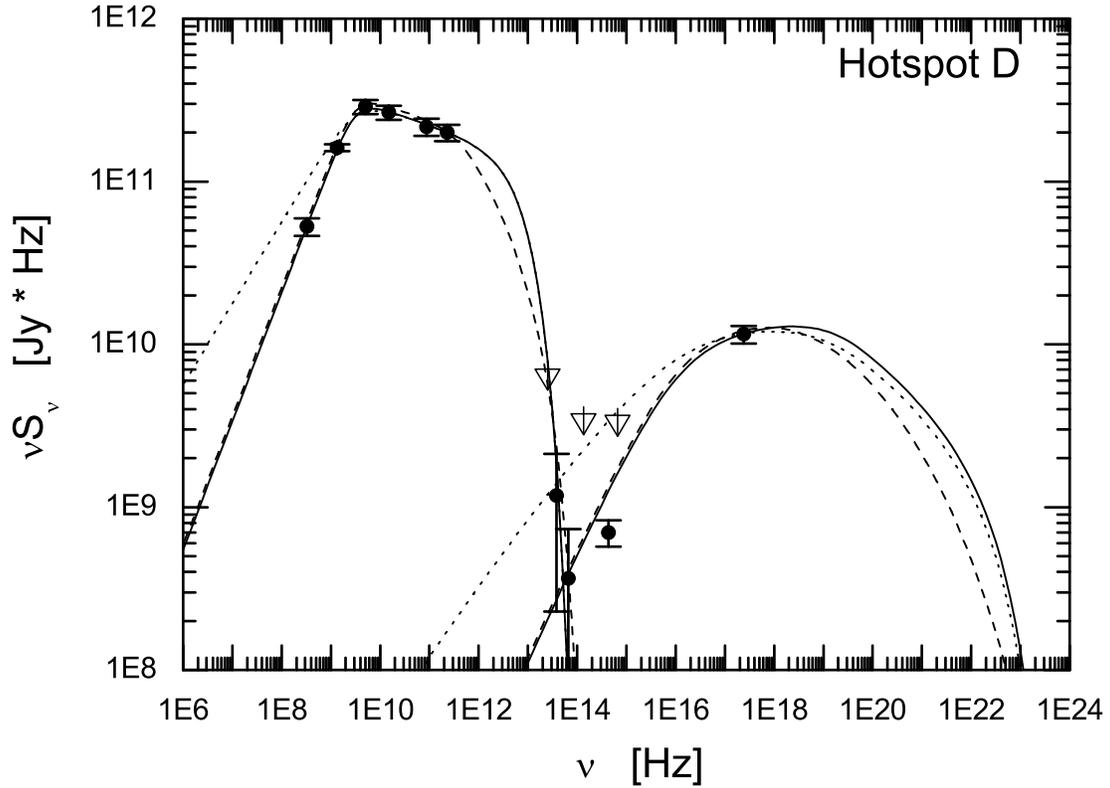}
\caption{Spectral energy distribution of hotspot D. Open triangles denote upper limits, while filled circles correspond to the detected fluxes. The parameters of the model described by the solid line are listed in Table~2. As discussed in \S~3.1, our modeling is little changed if we input an additional break ($\nu_{\rm br}=0.5 \times 10^{12}$\,Hz) in the synchrotron spectrum (dashed line). Increasing $\alpha_{\rm 1}$ to 0.5 (from 0.21) produces the dotted curves and overproduces the low-energy portion of the SSC component. All the lines assume $\gamma_{\rm min} = 1$.}
\end{figure}

\begin{figure}
\includegraphics[scale=1.50]{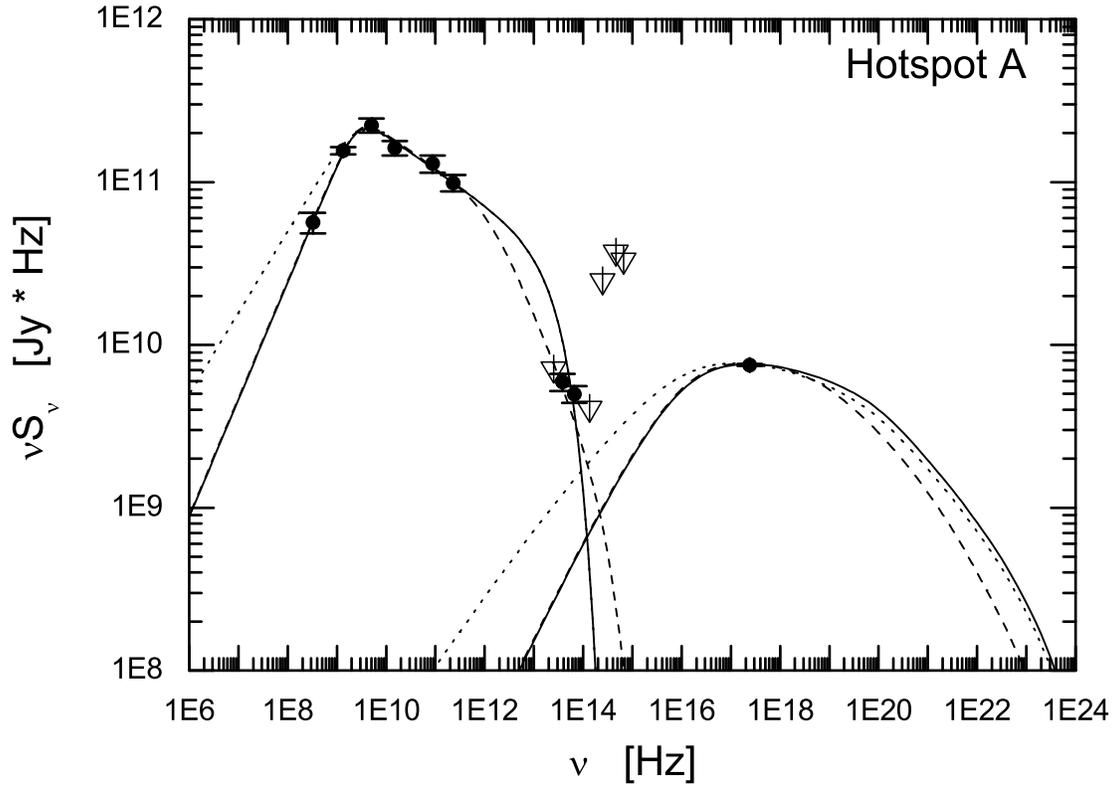}
\caption{Spectral energy distribution of hotspot A. Open triangles denote upper limits, while filled circles correspond to the detected fluxes. The parameters of the model described by the solid line are listed in Table~2. As in modeling the hotspot D (Figure~5), we show the effect of adding a break ($\nu_{\rm br}=1.2 \times 10^{12}$\,Hz) in the synchrotron spectrum (dashed line), or increasing $\alpha_{\rm 1}$ to 0.5 (dotted line). All the lines assume $\gamma_{\rm min} = 1$.}
\end{figure}

\begin{figure}
\includegraphics[scale=1.50]{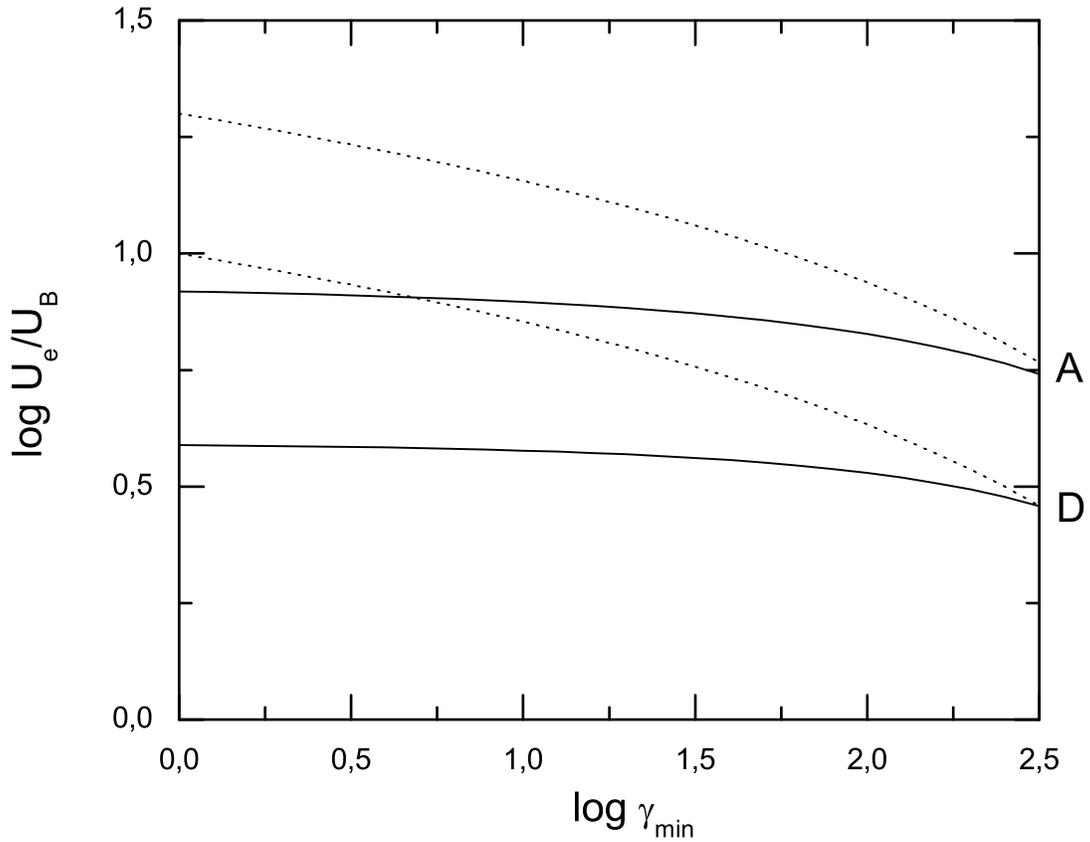}
\caption{Ratio of energy densities in ultrarelativistic electrons and magnetic field for hotspots A (lower solid and dotted lines) and D (upper solid and dotted lines). Solid lines correspond to the case with low-frequency synchrotron spectra $S_{\nu < \nu_{\rm cr}} \propto \nu^{-\alpha_1}$ with $\alpha_1 = 0.21$ (spot D) or $0.28$ (spot A), while dotted lines correspond to $\alpha_1 = 0.5$ for both features.}
\end{figure}

\begin{figure}
\includegraphics[scale=1.50]{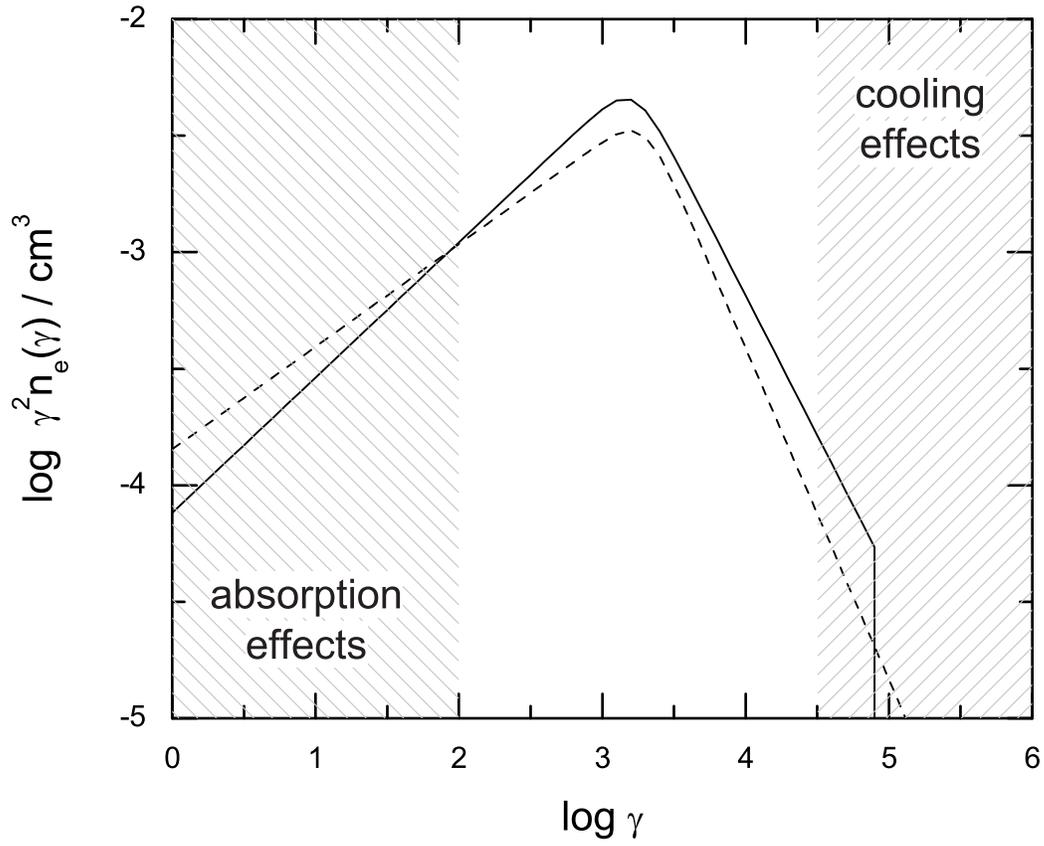}
\caption{Electron energy spectrum for hotspots A (dashed line) and D (solid line), as inferred from the SSC modeling. Shaded regions illustrates energy ranges affected by the absorption or radiative losses effects. Note the `normalizing factor' $\gamma^2$ in the abcissa scale.}
\end{figure}

\end{document}